# Моделирование эффективности многопереходных солнечных элементов

А.В. Саченко[+], В.П. Костылев[+], Н.Р. Кулиш[+], И.О. Соколовский[+], А.И. Шкребтий[*]

[+]Институт физики полупроводников им. В.Е. Лашкарева НАН Украины

03028 Киев, Украина (e-mail: sach@isp.kiev.ua)

[*]University of Ontario, Institute of Technology, Faculty of Science, 2000 Simcoe Street North, Oshawa, Ontario, Canada L1H 7K4

**Аннотация**

В настоящей работе при расчете эффективности $\eta$ многопереходных солнечных элементов (МПСЭ) учтены излучательная рекомбинация, рекомбинация Шокли – Рида, поверхностная рекомбинация на фронтальной и тыльной поверхностях, рекомбинация в областях пространственного заряда (ОПЗ), а также рекомбинация на границах гетеропереходов. Расчет выполнен путем самосогласованного решения уравнений для фототока и фотонапряжения, а также для теплового баланса. Учтено охлаждение МПСЭ по мере увеличения числа ячеек $n$ и улучшения условий теплоотвода. Рассмотрен эффект, приводящий к уменьшению фототока по мере увеличения $n$, связанный с сужением интервалов энергий фотонов, падающих на ячейку МПСЭ. Установлено, что существенное увеличение эффективности МПСЭ может быть достигнуто за счет улучшения условий теплоотвода, в частности, при использовании радиаторов и за счет приближения коэффициента серости МПСЭ к единице. Проведено сравнение полученных в работе результатов с результатами, приведенными в работах других авторов. Показано, что расчетные зависимости $\eta(n)$ согласуются со значениями, полученными экспериментально.

**Abstract**

The radiative recombination, Shokley-Read recombination, frontal-surface and rear-surface recombination and the recombination at the heterojunction boundaries and the recombination in the space charge region are considered in the calculation of the multijunction solar cell (MSC) efficiency. The calculation is performed by a self-consistent solution of the equations for the photocurrent and photovoltage, as well as the heat balance equation. A cooling of MSC with the increase of the number of cells $n$ and the improvement in the heat dissipation is regarded. It was found that, as the number of cells $n$ is increased, narrowing of spectral range for each cell causes additional reduction of current. A substantial increase in the MSC efficiency can be achieved by improving the heat extraction using radiators and increasing emissivity. A comparison is made between the calculated and experimental efficiency values. A rather good agreement was found. A comparison between this calculation and other formalisms is given.

# 1. Введение

В последние годы прилагаются большие усилия, направленные на разработку многопереходных солнечных элементов (МПСЭ) и увеличение их эффективности $\eta$. Так, в частности, в условиях АМ1.5 получены значения эффективности 37,8 %, а в условиях концентрированного освещения – 44,0 % [1]. Однако следует отметить, что в настоящее время существует разрыв между подходом ученых, разрабатывающих МПСЭ, и идеализированным подходом к моделированию их эффективности. Подавляющее большинство существующих расчетных работ выполнено, исходя из термодинамического подхода Карно или в модели энергетического баланса (см., например, [2-4]). Лишь одна известная нам работа оперирует с такими параметрами как токи насыщения для излучательной рекомбинации и рекомбинации Шокли-Рида [5]. Между тем для определения реальных значений эффективности МПСЭ и для оптимизации их параметров необходим более корректный подход, основанный на использовании физических закономерностей фотопреобразования в солнечных элементах и таких параметров полупроводниковых материалов как уровни легирования, толщины, времена жизни излучательной рекомбинации и рекомбинации Шокли – Рида, а также скорости поверхностной рекомбинации. Такой подход предложен в настоящей работе.

При моделировании эффективности учтен баланс между величиной $\eta(n)$ и температурой МПСЭ. Поскольку при большем числе ячеек $n$ уменьшается разница между ширинами запрещенных зон и энергиями фотонов, температура МПСЭ понижается. Вследствие понижения температуры возрастает напряжение разомкнутой цепи и соответственно увеличивается эффективность фотопреобразования. Особенно сильно данный эффект должен проявляться для солнечных элементов, работающих в космосе, поскольку в космических условиях температура не ограничена снизу такой высокой температурой, как в условиях Земли. Кроме того, показано, что величина температуры МПСЭ сильно зависит от условий теплоотвода.

Учтен также эффект, приводящий к уменьшению фототока по мере увеличения числа ячеек. Он проявляется из-за того, что в области края поглощения квантовый выход фототока $q_s(E_{ph})$ в зависимости от энергии фотонов $E_{ph}$ падает по сравнению с максимальной величиной, равной 1. При распределении солнечного спектра по большему числу ячеек площадь под кривой $q(E_{ph})$, равная фототоку в диапазоне энергий фотонов $\Delta E_{ph} = E_{ph2} - E_{ph1}$, уменьшается быстрее, чем поток фотонов в этом диапазоне. Поэтому в результате разбиения солнечного спектра на большее число интервалов фотонапряжение возрастает, а фототок дополнительно уменьшается. Это приводит к уменьшению эффективности фотопреобразования по мере увеличения количества ячеек и к насыщению зависимостей $\eta(n)$ при достаточно больших величинах $n$.

Показано, что улучшение теплообмена МПСЭ с окружающей средой, в частности, «зачернение» их тыловых поверхностей и использование радиаторов позволяют существенно увеличить эффективность МПСЭ.

Полученные в работе соотношения могут быть использованы для оптимизации параметров МПСЭ.

## 2. Постановка задачи

Выполнено моделирование эффективности фотопреобразования МПСЭ путем самосогласованного решения уравнений для фототока и фотонапряжения совместно с уравнением теплового баланса. Расчет произведен для СЭ, работающих при условиях AM0 и AM1.5. Рассмотрены две конструкции многопереходных СЭ, существующие в настоящее время: вертикальная и латеральная [6,4]. В первой конструкции $p-n$ переходы (или гетеропереходы), имеющие различные ширины запрещенных зон, располагаются друг за другом. При этом ближе к фотоприемной поверхности находятся более широкозонные, а затем – более узкозонные. В латеральной конструкции они располагаются в одной плоскости, однако на каждый $p-n$ переход (или гетеропереход) попадает лишь часть солнечного спектра. Его разложение производится с помощью дисперсионного элемента, например, дифракционной решетки.

В работе учтены следующие механизмы рекомбинационных потерь в МПСЭ: излучательная рекомбинация со временем жизни $\tau_r$, рекомбинация Шокли-Рида со временем жизни $\tau_{SR}$, рекомбинация в областях пространственного заряда (ОПЗ) $S_{sc}$, поверхностная рекомбинация на фронтальной и тыльной поверхностях $S_0$ и $S_d$, а также рекомбинация на пограничных состояниях в гетеропереходе $S_{get}$.

В случае, когда теплопроводность материалов отдельных ячеек и тепловой контакт между ними велики, так что в результате устанавливается одинаковая стационарная температура, для ее нахождения нужно использовать следующее уравнение теплового баланса

$$P_s(r-\eta(T)) = \beta K_T \sigma T^4. \tag{1}$$

Здесь $P_s = \int\limits_{E_1}^{E_2} P(E_{ph}) dE_{ph}$ – мощность падающего солнечного освещения, $P(E_{ph})$ – удельная мощность солнечного излучения для данной энергии фотона, $r \leq 1$, $E_1$ и $E_2$ – граничные энергии фотонов, задающие энергетический интервал, $\eta(T)$ – эффективность МПСЭ, $\beta$ – численный коэффициент порядка 1. Он зависит от коэффициента серости, т.е. от близости излучения СЭ к излучению абсолютно черного тела. Величина $K_T$ равна отношению площади излучающей поверхности к площади освещаемой поверхности. Параметр $\sigma$ – постоянная Стефана – Больцмана. Величина $T = T_{min} + \Delta T$, где $T_{min}$ – температура окружающей среды, $\Delta T$ – величина превышения полной температуры $T$ по сравнению со значением $T_{min}$.

Коэффициент $r$, меньший 1, учитывает, что неполное поглощение света в МПСЭ не приводит к повышению температуры, а также учитывает то, что излучательная рекомбинация уносит часть энергии солнечного излучения. Это также не сопровождается повышением температуры МПСЭ.

При нахождении температуры СЭ для условий AM1.5, для простоты, также предполагается, что и в данном случае действует лишь радиационный механизм охлаждения. Здесь и далее будем считать, что МПСЭ имеет единичную площадь. При этом удельная плотность фототока и фототок численно равны друг другу.

Остановимся более подробно на анализе вопроса о значениях минимальных температур солнечных батарей при условиях освещения AM1.5 и AM0. В условиях AM1.5 минимальная температура равна температуре окружающей среды. В условиях AM0 нагрев СЭ происходит как благодаря превращению в теплоту энергии части солнечного спектра, так и из-за поглощения солнечной батареей теплового излучения спутника и отраженного от спутника солнечного освещения. Кроме того, разогрев космического аппарата (и солнечной батареи) происходит в результате трения с остаточной атмосферой. Влиянием поглощения солнечной батареей теплового излучения спутника и отраженного от спутника солнечного излучения на величину температуры СЭ можно пренебречь, выбрав их соответствующую взаимную ориентацию. Что касается трения, то соответствующий разогрев будет сильно зависеть от орбиты спутника и ориентации солнечной батареи по отношению к направлению движения. Так, например, для случая низких орбит величина $T_{min}$ будет больше, чем для высоких. Как показано в работе [7], даже в случае низких орбит спутника величина $T_{min}$ составляет около –100° С. Поэтому в дальнейших расчетах величина $T_{min}$ в условиях AM0 полагается равной 173 К.

Перейдем далее к выводу соотношений, позволяющих определить эффективность МПСЭ в зависимости от ширин запрещенных зон используемых полупроводников.

Величина напряжения разомкнутой цепи $V_{oc}$ одной ячейки МПСЭ с учетом рекомбинации в ОПЗ $S_{SC}$ и величины суммарной скорости поверхностной рекомбинации $S_s$ находится из такого уравнения:

$$J_g = q\left(\frac{d}{\tau}+S_s\right)\frac{n_i^2}{n_0}\left[\exp\left(\frac{qV_{oc}}{kT}\right)-1\right] + q\frac{\kappa L_D n_i}{\tau_{SR}\sqrt{\frac{1}{2}\ln\left(\frac{qn_0(d/\tau+S_s)}{J_g}\right)}}\left[\exp\left(\frac{qV_{oc}}{2kT}\right)-1\right]. \qquad (2)$$

Здесь $J_g$ – плотность протекающего фототока, $q$ – элементарный заряд, $d$ – толщина, $n_0$ – равновесная концентрация основных носителей заряда, $\tau = \left(\tau_r^{-1}+\tau_{SR}^{-1}\right)^{-1}$ – объемное время жизни, $\tau_r = (An_0)^{-1}$ – излучательное время жизни, $A$ – параметр излучательной рекомбинации, $\tau_{SR}$ – время жизни Шокли-Рида, $\kappa \approx 2$, $L_D = \left(\varepsilon_0\varepsilon_s kT/2q^2 n_0\right)^{1/2}$ – дебаевская длина экранирования, $\varepsilon_0$ – диэлектрическая константа вакуума, $\varepsilon_s$ – относительная диэлектрическая проницаемость полупроводника. $k$ – постоянная Больцмана,

$$n_i(T) = \sqrt{N_c N_v}\left(\frac{T}{300}\right)^{3/2}\exp\left(-\frac{E_g}{2kT}\right) \qquad (3)$$

– концентрация носителей заряда в собственном полупроводнике, $N_c$ и $N_v$ – эффективные плотности состояний зоны проводимости и валентной зоны при T =300 К, $E_g$ – ширина запрещенной зоны.

Уравнение (2) справедливо, если выполнено неравенство $L > d$, где $L = (D\tau)^{1/2}$ и $D$ соответственно длина и коэффициент диффузии неосновных носителей заряда. Первое слагаемое в правой части (2) записано в обычной форме [8]. Второе слагаемое, описывающее вклад рекомбинации в ОПЗ, модифицировано нами по сравнению с [8] для случая освещения.

С другой стороны, уравнение (2) для $i$-той ячейки МПСЭ можно переписать в таком виде:

$$J_{gi} = J_{0di}\exp\left(\frac{qV_{oci}}{kT}\right) + J_{0ri}\exp\left(\frac{qV_{oci}}{2kT}\right) = J_{0i}^*\exp\left(\frac{qV_{oci}}{m_i kT}\right). \qquad (4)$$

Здесь $J_{0di}$ – плотность диффузионного тока насыщения, $J_{0ri}$ – плотность рекомбинационного тока насыщения, $J_{0i}^*$ – плотность эффективного тока насыщения, а $m_i$ – эффективная величина коэффициента идеальности ВАХ для $i$-той ячейки.

Сопоставляя (2) и (4), можно определить величины $J_{0di}$ и $J_{0ri}$ с учетом рассматриваемых рекомбинационных механизмов.

Согласно работе [9] для случая $n$ соединенных последовательно ячеек напряжение разомкнутой цепи МПСЭ $V_s$ определяется следующими соотношениями

$$V_s = \frac{kT}{q}\sum_{i=1}^{n}\ln\left(\frac{J_g}{J_{0i}^*}\right)^{m_i} = m_s\frac{kT}{q}\ln\left(\frac{J_g}{J_{0s}}\right), \qquad (5)$$

где

$$m_s = m_1 + m_2 + ... m_n, \qquad (6)$$

$$J_{0s} = \left(J_{01}^{*m_1} \cdot J_{02}^{*m_2} \cdot ... \cdot J_{0n}^{*m_n}\right)^{\frac{1}{m_s}}. \qquad (7)$$

Вольт-амперная характеристика для последовательно соединенных $n$ ячеек в данном случае определяется уравнением

$$J(V) = J_g - J_{0s}\exp\left(\frac{qV}{m_s kT}\right), \qquad (8)$$

где $V$ – приложенное напряжение.

Используя условие

$$\frac{d}{dV}(J(V)V) = 0, \qquad (9)$$

можно получить трансцендентное уравнение для определения фотонапряжения в точке максимального отбора мощности $V_m$ вида

$$V_m \cong V_s\left(1 - \frac{m_s kT\ln(qV_m/m_s kT)}{qV_m}\right). \qquad (10)$$

Отметим, что для МПСЭ обычно выполняется условие $qV_m/m_s kT >> 1$, поэтому уравнение (10) может быть решено методом последовательных приближений. Зачастую уже первого приближения достаточно для того, чтобы найти величину $V_m$ с заданной точностью.

Плотность тока короткого замыкания может быть найдена как для латеральных, так и для вертикальных МПСЭ. Так, в случае латеральных МПСЭ справедливо соотношение

$$J_{gi}^L(E_{gi}) = s_i^{-1}\int_{E_1}^{E_2}j_{gi}(E_{ph})q_{si}(E_{gi}, E_{ph})dE_{ph}, \qquad (11)$$

где $s_i = S_i/S$ – удельная площадь $i$-ой ячейки, $j_g(E_{ph})$ – удельная плотность фототока для энергии фотона $E_{ph}$, определяемая произведением элементарного заряда $q$ на плотность потока солнечного излучения, $q_{si}(E_{gi}, E_{ph})$ – квантовый выход фототока.

Достаточно общее выражение для величины $q_{si}(E_{gi}, E_{ph})$ приведено, например, в [10]. Оно имеет следующий вид

$$q_s = q_{sp} + q_{sn}, \qquad (12)$$

$$q_{sp} = \frac{\alpha L_p}{(\alpha L_p)^2 - 1} \cdot \frac{\alpha L_p + S_0 \frac{\tau_p}{L_p}\left(1 - e^{-\alpha d_p}\right)\cosh\left(\frac{d_p}{L_p}\right) - e^{-\alpha d_p}\sinh\left(\frac{d_p}{L_p}\right) - \alpha L_p e^{-\alpha d_p}}{S_0 \frac{\tau_p}{L_p}\sinh\left(\frac{d_p}{L_p}\right) + \cosh\left(\frac{d_p}{L_p}\right)}, \qquad (13)$$

$$q_{sn} = \frac{\alpha L e^{-\alpha d_p}}{1 - (\alpha L)^2} \cdot \left\{ \frac{\left[S_d \cosh\left(\frac{d}{L}\right) + \frac{D}{L}\sinh\left(\frac{d}{L}\right)\right]\left(1 + R_d e^{-2\alpha d}\right) + (\alpha D(1 - R_d) - S_d(1 + R_d))e^{-\alpha d} -}{S_d \sinh\left(\frac{d}{L}\right) + \frac{D}{L}\cosh\left(\frac{d}{L}\right)} \right.$$

$$\left. \frac{-\alpha L S_d\left[\sinh\left(\frac{d}{L}\right) + \frac{D}{L}\cosh\left(\frac{d}{L}\right)\right]\left(1 - R_d e^{-2\alpha d}\right)}{S_d \sinh\left(\frac{d}{L}\right) + \frac{D}{L}\cosh\left(\frac{d}{L}\right)} \right\}. \qquad (14)$$

Здесь $\alpha$ – коэффициент поглощения света, $L_p$ – длина диффузии в эмиттере, $d_p$ – толщина эмиттера, $\tau_p$ – объемное время жизни в эмиттере, $R_d$ – коэффициент отражения света от тыльной поверхности.

Отметим, что ранее было принято, что всегда выполняется неравенство $L > d$. С его учетом выражение для $q_{sn}$ может быть упрощено, однако более существенным является то, что в данном случае нет необходимости отдельно учитывать квантовый выход фототока в ОПЗ [8]. Фактически это нужно делать лишь тогда, когда для длины диффузии $L$ выполняется соотношение $L \leq w$, где $w$ – толщина ОПЗ.

В случае вертикальных МПСЭ выражение для плотности тока короткого замыкания $J_{gi}^V$ имеет вид

$$J_{gi}^V(E_{gi}) = \int_{E_1}^{E_2} j_g(E_{ph}) q_{si}(E_{gi}, E_{ph}) T_i(E_{gi}, E_{ph}) dE_{ph}, \qquad (15)$$

где

$$T_i(E_{gi}, E_{ph}) = e^{-\alpha_1 d_1 - \alpha_2 d_2 \cdots -\alpha_{i-1} d_{i-1}}. \qquad (16)$$

Для квантового выхода $i$-той ячейки в этом случае также можно использовать выражения (12) –(14) с нулевым коэффициентом отражения $R_{di}$.

При последовательном соединении ячеек МПСЭ необходимо согласование по току, т.е.

$$J_{g1}(E_{g1}) = J_{g2}(E_{g2}) = \cdots = J_{gn}(E_{gn}) = J_g. \qquad (17)$$

Выражение для плотности фототока в точке отбора максимальной мощности $J_m$, как обычно, определяется соотношением

$$J_m = J_g\left(1 - \frac{r_s kT}{qV_m}\right), \qquad (18)$$

а эффективность МПСЭ с учетом только рассмотренных выше рекомбинационных потерь равна

$$\eta_0 = \frac{J_m V_m}{P_s}. \qquad (19)$$

Для случая МПСЭ она может быть найдена совместным решением уравнений и соотношений (1) – (19).

Окончательно выражение для эффективности МПСЭ с учетом всех потерь может быть записано в виде

$$\eta = Q_{dr}(1 - R_s)(1 - K_m)\eta_0. \qquad (20)$$

Здесь $Q_{dr}$ -эффективность дисперсионного элемента, $R_s$ – коэффициент отражения света от фронтальной поверхности МПСЭ, а $K_m$ – коэффициент ее затенения.

Отдельно следует рассмотреть случай, когда ячейки МПСЭ формируются на основе гетеропереходов. При этом, как правило, в качестве окна используется более широкозонный полупроводник, чем поглотитель (база). Это, как показано, в частности, в работе [11], приводит к тому, что скорость поверхностной рекомбинации на фронтальной (освещенной) поверхности $S_0$ становится малой по сравнению со скоростью рекомбинации в объеме. Однако в случае достаточно сильного различия постоянных решеток используемых полупроводников возникает большая концентрация пограничных состояний и, соответственно, большая скорость рекомбинации на этих состояниях $S_{get}$.

Качественным отличием рекомбинации, происходящей на гетеропереходе, от рекомбинации на поверхности МПСЭ, является то, что она, как правило, не влияет на ток короткого замыкания и проявляется только в режиме разомкнутой цепи. Это связано с тем, что анизотипный гетеропереход расположен в области сильного электрического поля, которое эффективно разводит генерируемые светом электроны и дырки. Поэтому в случае гетероперехода не нужно учитывать рекомбинацию на освещенной поверхности, а рекомбинацию в гетеропереходе необходимо рассматривать только при расчете напряжения разомкнутой цепи $V_{oc}$.

Конкретные расчеты были проведены нами для случая прямозонных полупроводников группы $A_3B_5$.

Прежде чем перейти к изложению и анализу полученных результатов, рассмотрим предельный случай, когда единственным рекомбинационным механизмом является излучательная рекомбинация. В этом случае из уравнения (2) в явном виде может быть найдена величина напряжения разомкнутой цепи $i$-той ячейки МПСЭ. Запишем ее в таком виде

$$V_i = E_{gi} - \frac{kT}{q}\ln(1+\nu_i) = E_{gi} - V_{ri}, \qquad (21)$$

где

$$\nu_i = \frac{qd_i A_i N_{ci} N_{vi}\left(\dfrac{T}{300}\right)^3}{J_g}. \qquad (22)$$

Как видно из выражений (21) и (22), напряжение разомкнутой цепи $i$-той ячейки МПСЭ $V_i$ в данном случае состоит из двух членов. Первый член – ширина запрещенной зоны $E_{gi}$, а второй, $V_{ri}$, практически всегда, за исключением случая малых значений $E_{gi} \le 0.5$ эВ, линейно зависит от температуры. Так, при типичных параметрах прямозонных полупроводников группы $A_3B_5$, когда $A \approx 2\cdot10^{-10}$ см$^3$/с, $N_c \approx 5\cdot10^{17}$ см$^{-3}$, $N_v \approx 10^{19}$ см$^{-3}$ [12], $d \approx 1$ мкм, T=300 К, $J_g \approx 10$ мА/см$^2$ величина $\nu \approx 10^5$, а $V_r \approx 0.3$ В. Поскольку в данном случае аргумент логарифма существенно больше 1, то сравнительно небольшая разница в значениях A, $N_c$ и $N_v$, имеющая место для различных полупроводников [12], не приведет к значительному изменению величины $V_r$. Поэтому для приближенного нахождения значения $\eta(n)$ в случае полупроводников с произвольными ширинами запрещенных зон можно воспользоваться одинаковыми величинами $\nu_i$.

Кроме того, поскольку информация о величинах коэффициентов поглощения $\alpha_i$ и их зависимостях от энергии фотона для достаточно большого количества прямозонных полупроводников группы $A_3B_5$ отсутствует, мы, чтобы получить зависимости $\alpha(E_g, E_{ph})$ от энергии фотона для произвольного значения $E_g$, воспользуемся следующим приемом.

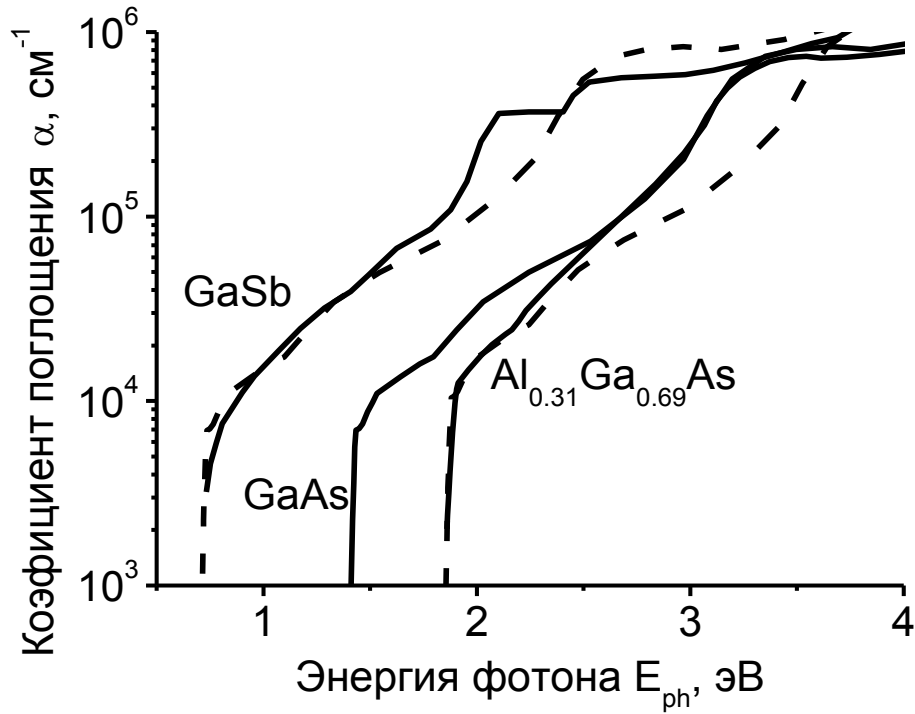

Рис.1. Зависимости коэффициента поглощения света от энергии фотонов для ряда прямозонных полупроводников.

На рис. 1 приведены экспериментальные зависимости $\alpha(E_g, E_{ph})$ для трех реперных полупроводников: GaSb, GaAs и $Al_{0.45}Ga_{0.55}As$ [12]. Смещая зависимости $\alpha(E_g, E_f)$ для арсенида галлия по энергии в сторону уменьшения величины $E_g$ и совместив их с экспериментальными зависимостями для GaSb, можно получить достаточно хорошее совпадение между ними.

Аналогично смещение зависимости $\alpha(E_g, E_{ph})$ для GaAs в сторону больших величин $E_g$ позволяет совместить ее с хорошей точностью с экспериментальной зависимостью для случая $Al_{0.31}Ga_{0.69}As$ (рис. 1). И в одном, и в другом случае речь идет о хорошем совмещении в области, где $\alpha(E_g, E_{ph})d_i \leq 1$. В области $\alpha(E_g, E_{ph})d_i \gg 1$ совпадение отсутствует, однако оно здесь и не нужно, поскольку квантовый выход фототока в данном случае равен 1, а поглощение света полное. Поэтому расчеты, использующие как реперные, так и совмещенные зависимости $\alpha(E_g, E_{ph})$, дадут правильный результат. Это дало возможность использовать в дальнейшем экстраполированные таким образом исходные зависимости для GaAs в случае произвольных значений $E_{gi}$.

### 3. Полученные результаты и их анализ

В данной работе все расчеты выполнены для случая, когда СЭ поглощает солнечный спектр, лежащий в диапазоне длин волн $0.3 \mu m < \lambda < 2 \mu m$. Нефотоактивная часть солнечного спектра с $\lambda > 2 \mu m$ не попадает на СЭ. Это способствует понижению температуры МПСЭ.

Для простоты при дальнейших расчетах мы ограничивались лишь учетом рекомбинационных потерь. При этом выполняется условие $\eta(n) = \eta_0(n)$.

При построении конкретных зависимостей $\eta(n)$ в случае произвольных величин $E_{gi}$ были использованы следующие параметры: $A = 2 \cdot 10^{-10}$ см$^3$/с, $D = 50$ см$^3$/с, $N_c = 5 \cdot 10^{17}$, $N_v = 10^{19}$ см$^{-3}$. Полагалось также, что $d = 2 \cdot 10^{-4}$ см, $d_p = 10^{-6}$ см, $\tau_p = 10^{-10}$ с, $n_0 = 10^{17}$ см$^{-3}$. Считалось, что значения скоростей поверхностной рекомбинации и времен жизни Шокли-Рида одинаковы для всех ячеек МПСЭ. При учете рекомбинации Шокли-Рида $\tau_{SR}$ полагалось равным $5 \cdot 10^{-9}$ с.

В работе [5] для получения больших значений эффективности $\eta(n)$ учитывались процессы двукратного или многократного отражения света и перепоглощения. В настоящей работе рост эффективности МПСЭ достигается, в основном, за счет улучшения теплоотвода. Одной из таких возможностей является увеличение значения $\beta$, что соответствует приближению характера теплового излучения МПСЭ к излучению абсолютно черного тела. Так, например, известно, что величина $\beta$ для случая, когда тепловое излучение происходит только через освещаемую поверхность СЭ, близка к 0.9 [13]. По-видимому, «зачернением» тыловой поверхности СЭ можно довести значение $\beta$ до величины, близкой к 1.8.

Другим способом улучшения теплообмена является использование радиаторов, т.е. реализации ситуации, когда тепловое излучение происходит с больших площадей, чем площадь освещаемой поверхности МПСЭ ($K_T > 1$).

На рис. 2 приведены зависимости $\eta(n)$ для случая произвольных значений $E_{gi}$ в условиях АМ0 (рис. 2а) и АМ1.5 (рис. 2б) при пренебрежении всеми каналами рекомбинации, кроме излучательного. Расчет был выполнен для латеральной конструкции МПСЭ. Наибольшие значения эффективности $\eta(n)$ отвечают предельно возможным значениям (кривые 1). При получении предельных величин $\eta(n)$ считалось, что напряжения разомкнутой цепи всех ячеек равны величинам $E_{gi}$, а значения токов

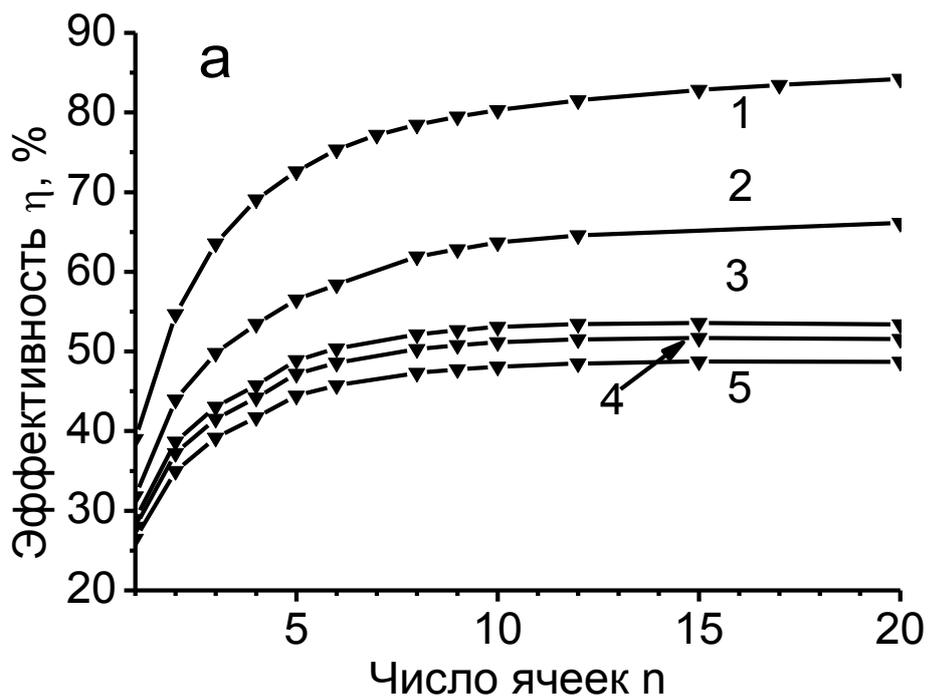

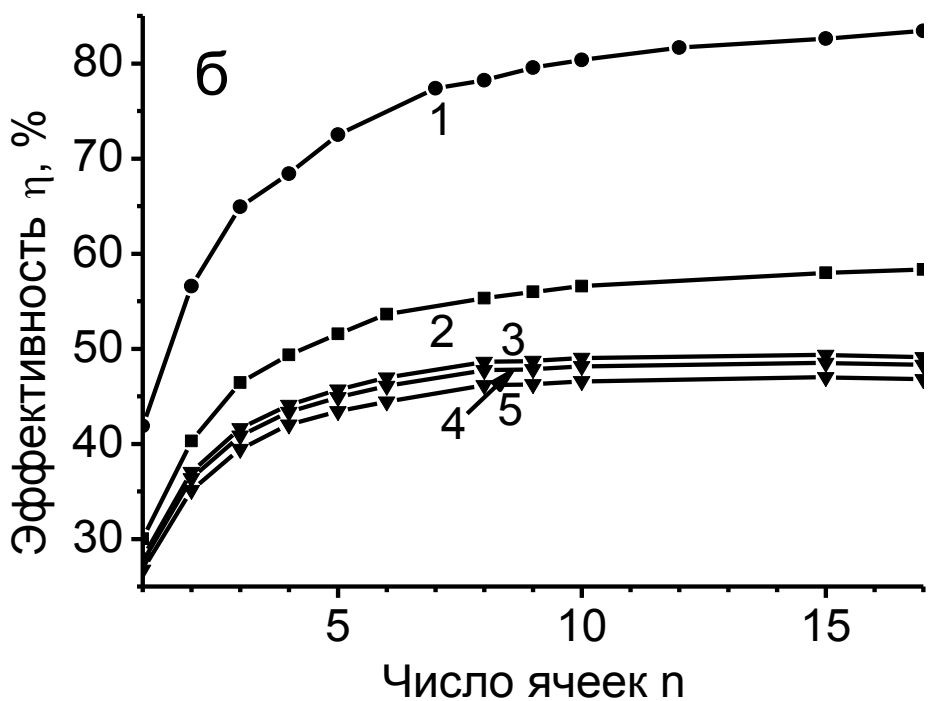

Рис. 2. Зависимости эффективности $\eta(n)$ для произвольного набора полупроводников в условиях AM0 (рис. 2а) и в условиях AM1.5 (рис. 2б) при доминировании излучательной рекомбинации.

короткого замыкания определяются лишь энергетическим интервалом солнечного излучения, попадающего на $i$-тую ячейку. Фактор заполнения ВАХ определялся из условия, что предельная температура равна 173 К в условиях АМ0 и 300 К в условиях АМ1.5. Как видно из рис. 2, получаемые при этом величины предельных значений при $n=16$ очень большие и превышают 80 %. Естественно, что эти значения не могут быть реализованы экспериментально. Однако в отличие от подходов, описанных в [2-4], они получены при использовании фотоэлектрического подхода.

Наибольшие достигаемые значения эффективности $\eta(n)$ в рамках развитого выше подхода получаются в пренебрежении всеми каналами рекомбинации, кроме излучательного, при использовании следующих условий теплоотвода: $\beta=1.5$, $K_T=5$, а также для случая, когда величина $R_d=0.8$ (кривые 2). Как видно из рисунка 2, в этом случае также достигаются большие значения $\eta(n)$, особенно в условиях АМ0. Однако эти значения эффективности МПСЭ, в принципе, могут быть достигнуты экспериментально, если реализовать указанные выше условия теплоотвода и свести к минимуму неизлучательные каналы рекомбинации.

Если $R_d=0$, а $K_T=1$, то регулировать условия теплоотвода можно лишь за счет изменения $\beta$. На кривых 3-5, приведенных на рис. 2, расчет эффективности $\eta(n)$ был выполнен при использовании значений $\beta$, равных 2, 1.5 и 1. Как видно из рис. 2, с увеличением $\beta$ величина эффективности возрастает.

Отметим, что чем выше эффективность $\eta(n)$, тем ниже температура. Так, в условиях АМ0 температура изменяется для кривых 1-5 в диапазоне от 173 до 320 К, а в условиях АМ1.5 – от 300 до 360 К.

На рис. 3 приведены зависимости $\eta(n)$ для случая произвольных значений $E_{gi}$ в условиях АМ0 (рис. 3а) и АМ1.5 (рис. 3б) при учете излучательной рекомбинации и рекомбинации Шокли-Рида со временем жизни $5\cdot10^{-9}$ с. Естественно, что в этом случае актуальны как рекомбинация Шокли-Рида в нейтральной области базы, так и рекомбинация в ОПЗ.

Из сравнения рис. 2 и 3 видно, что включение дополнительного канала рекомбинации Шокли-Рида с меньшим значением времени жизни, чем время излучательной рекомбинации, приводит к уменьшению эффективности $\eta(n)$. Вместе с тем, в условиях АМ1.5, а особенно в условиях АМ0 максимальные значения эффективности при $n=16$ остаются достаточно большими.

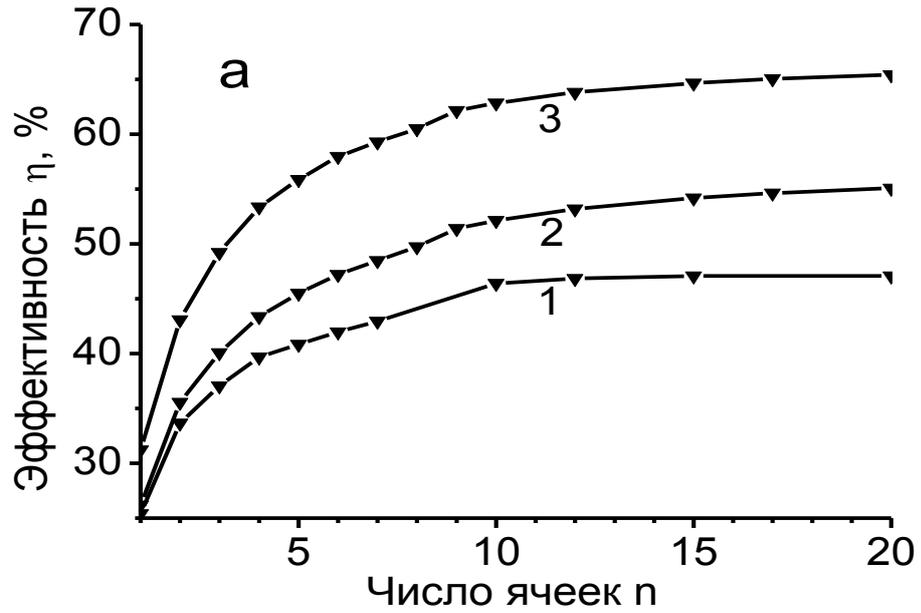

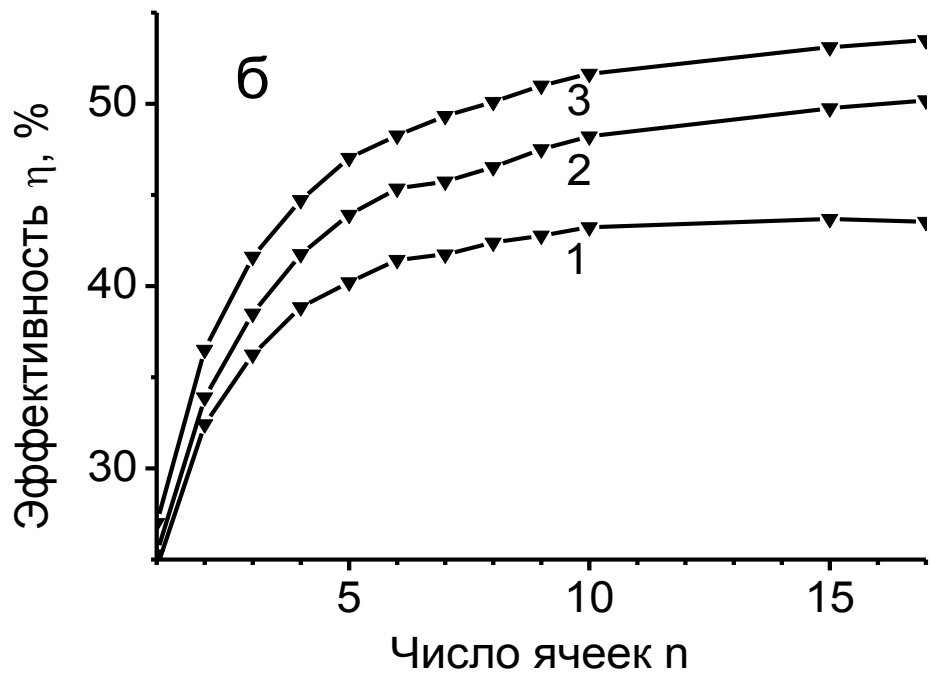

Рис. 3. Зависимости эффективности $\eta(n)$ для произвольного набора полупроводников в условиях АМ0 (рис. 3а) и в условиях АМ1.5 (рис. 3б) при доминировании рекомбинации Шокли-Рида.

На рис. 4 приведены расчетные зависимости величины $\eta(n)$ в условиях АМ1.5 для конкретного набора полупроводников (GaSb, GaAs$_{0,7}$Sb$_{0,3}$, GaAs, Al$_{0,15}$Ga$_{0,85}$As, Al$_{0,3}$Ga$_{0,7}$As, Al$_{0,45}$Ga$_{0,55}$As) при учете $\tau_r$ и рекомбинации Шокли-Рида с временем жизни $5 \cdot 10^{-9}$ с. При построении зависимостей $\eta(n)$ также модифицировались условия теплоотвода. Как видно из рисунка, в данном случае максимальное значение эффективности достигается при $n=4$. Сильное уменьшение эффективности при больших значениях $n$ и реализация максимума на зависимостях $\eta(n)$ при $n=4$ связаны с неидеальным совпадением ширин запрещенных зон для выбранных полупроводников по сравнению с идеальным случаем, рассмотренным ранее.

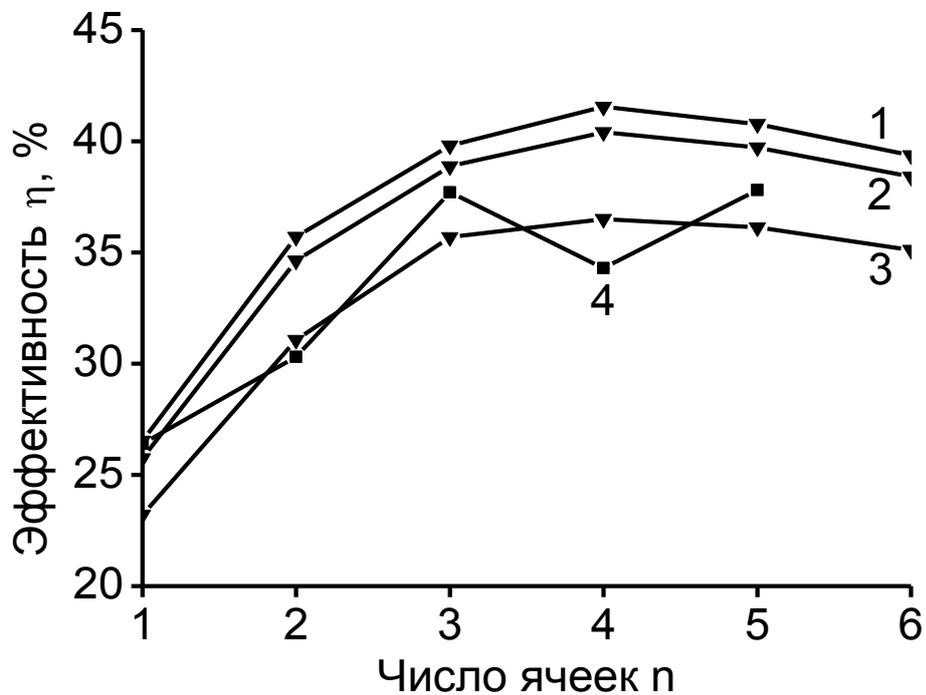

Рис. 4. Зависимости эффективности $\eta(n)$ для конкретного набора полупроводников в условиях АМ1.5 при доминировании рекомбинации Шокли-Рида.

При построении кривых 1 – 3 считалось, что $R_d = 0$, а $\beta =1.5$. Наименьшее значение эффективности получено в условиях, когда $K_T = 1$ (кривая 3). Большая величина эффективности получается при условии, когда $K_T =5$ (кривая 2). В условиях принудительного охлаждения, когда температура МПСЭ равна 300 К, получается наибольшее значение $\eta(n)$ (кривая 1). Величина эффективности в максимуме при изменении условий теплоотвода возрастает.

Сравнивая между собой результаты, полученные в условиях AM1.5 для произвольных по ширинам зон полупроводников (рис. 3б) и для набора реальных полупроводников (рис. 4), можно увидеть что максимальные значения эффективности $\eta(n)$ в первом случае примерно на 10 % выше, чем во втором. Отсюда можно сделать вывод о том, что существует достаточно большой резерв в повышении эффективности МПСЭ в условиях AM1.5 за счет более точного согласования требуемых ширин запрещенных зон прямозонных полупроводников $E_{gi}$ по мере увеличения $n$. Использовать этот резерв можно за счет синтеза прямозонных полупроводников с необходимыми ширинами запрещенных зон и требуемыми значениями других параметров.

На кривой 4 приведены экспериментальные зависимости $\eta(n)$, полученные в работах [14 - 16]. Как видно из сравнения теоретических зависимостей с экспериментальными значениями, при ухудшении условий теплоотвода реализуются даже меньшие значения эффективности, чем полученные экспериментально. Так, из сравнения кривых 3 и 4, видно, что экспериментальные значения для $n$ =1, 3 и 5 больше, чем полученные для теоретической кривой 3.

Естественно, что из этого не следует делать особых выводов, поскольку различны как комбинации полупроводников, использованные при расчете и в эксперименте, так и условия минимизации в эксперименте других потерь, связанных, в частности, с поверхностной рекомбинацией. Однако, с учетом того, что механизмы потерь в прямозонных полупроводниках группы $A_3B_5$, использованных как для расчета, так и в эксперименте, достаточно схожи, неплохое соответствие между теоретическими и экспериментальными зависимостями $\eta(n)$ следует признать не случайным.

Подобные расчеты можно провести и для МПСЭ вертикальной конструкции. Однако в этом случае, как отмечалось выше, нельзя таким образом вводить для каждой отдельной ячейки коэффициент отражения от тыловой поверхности $R_d$. Более корректно положить его равным нулю. В связи с этим, величины $\eta(n)$ для МПСЭ вертикальной конструкции при условиях $\beta$ =2 и $K_T$ =5 будут меньше, чем полученные для МПСЭ латеральной конструкции при $R_d$ =0.8, $\beta$ =2 и $K_T$ =5.

Возникает вопрос, как повлияет поверхностная рекомбинация на величину $\eta(n)$? При использованных выше значениях $d$ =2·10$^{-4}$ см и $\tau_{SR}$ =5·10$^{-9}$ с скорость рекомбинации в объеме полупроводника составляет 4·10$^4$ см/с. Это означает, что скорость поверхностной рекомбинации приведет к падению эффективности, если $S_s$ порядка или больше 4·10⁴ см/с. Как показано в работах Ж.И. Алферова с соавторами, величины скоростей поверхностной рекомбинации (или скоростей рекомбинации на гетеропереходе) в полупроводниках группы $A_3B_5$ составляют значения $\leq 10^4$ см/с для случая хорошо согласованных по значениям постоянных решетки полупроводников, например, в системе GaAs-AlGaAs (см. также [17]). Если же хорошего согласования нет, то типичные значения $S_s \sim 10^5$ см/с. Поэтому, если учесть, что в случае, когда $\tau_{SR}$=5·10⁻⁹ с, кроме объемной рекомбинации существенна и рекомбинация в ОПЗ, то наличие рекомбинации Шокли-Рида с приведенным значением времени жизни дает примерно такое же уменьшение эффективности, как и $S_s \approx 10^5$ см/с.

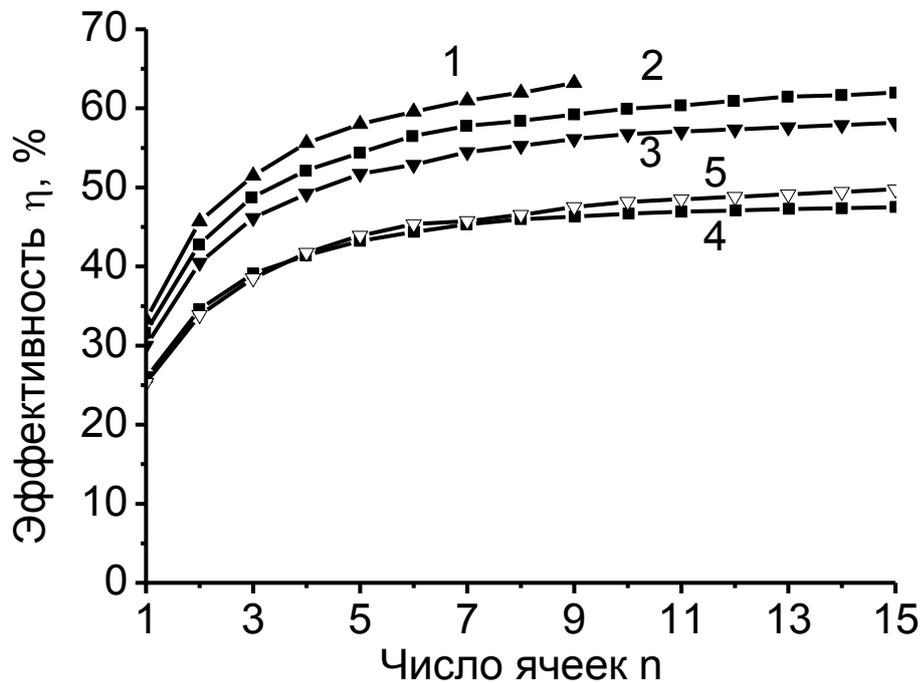

Рис. 5. Расчетные зависимости эффективности $\eta(n)$ для произвольного набора полупроводников в условиях AM1.5, полученные в работах [5], [18] и в настоящей работе.

На рисунке 5 приведены, во-первых, три расчетные зависимости для предельно достижимых значений эффективности МПСЭ $\eta$ от числа ячеек $n$. Верхняя кривая (1) взята из работы [18]. Эта зависимость получена в модели энергетического баланса. Кривая 2 соответствует верхней зависимости $\eta(n)$, приведенной в работе [5] для случая, когда единственным рекомбинационным механизмом является излучательная рекомбинация (см. рис. 19). Кривая 3 соответствует полученной в нашей работе зависимости, описываемой кривой 2, приведенной на рис. 2б. Она также построена в пренебрежении всеми механизмами рекомбинации, кроме излучательной, для условий эффективного теплоотвода. Все три кривые описывают предельно достижимые эффективности МПСЭ. Различие полученных значений $\eta(n)$ на различных кривых составляет ≤ 10 %.

На кривых 4 и 5, также приведенных на этом рисунке, построены зависимости $\eta(n)$, полученные в работе [5], а также в нашей работе, для случая, когда преобладающим механизмом рекомбинации является рекомбинация Шокли – Рида. Кривая 4 взята из рисунка 19 работы [5] для случая, когда плотность тока насыщения для механизма рекомбинации Шокли-Рида равна 100 А/см$^2$, а кривая 5 соответствует кривой 3, приведенной на рис. 2б. Как видно из их сравнения, приведенные зависимости $\eta(n)$ практически совпадают между собой. Это не случайное совпадение, поскольку расчет плотностей токов насыщения излучательной рекомбинации и рекомбинации Шокли – Рида, выполненный с использованием формулы (2) при подстановке соответствующих параметров, дает близкие значения по сравнению с приведенными в [5] плотностями соответствующих токов насыщения.

Мы не будем анализировать детально в данной работе вопрос о значениях $\eta(n)$ в условиях концентрированного освещения. Отметим только, что для роста эффективности МПСЭ в этом случае необходимо, чтобы степень концентрации освещения М была пропорциональной величине $K_T$. В противном случае с ростом М произойдет разогрев МПСЭ и эффективность упадет. Естественно, что достаточно большие значения $K_T$ можно реализовать только в условиях АМ1.5. В космических условиях сделать это затрудняет ограничение веса МПСЭ.

## 4. Выводы

В настоящей работе предложен подход к расчету реальных значений для эффективности многопереходных солнечных элементов. Показано, что при учете объемной рекомбинации Шокли-Рида, а также поверхностной рекомбинации со скоростями $S_0$, $S_d$ и $S_{get}$ величина $\eta(n)$ уменьшается по сравнению со случаем, когда доминирует излучательная рекомбинация.

В то же время в случае произвольных по ширинам запрещенных зон полупроводников расчетные значения максимальной эффективности МПСЭ при улучшенных условиях теплоотвода достигают больше 60 % для условий АМ0 и больше 50 % для условий АМ1.5.

Показано, что максимум величины $\eta(n)$ для выбранных нами комбинаций полупроводниковых материалов в условиях AM1.5 достигается при $n=4$. При использованных значениях параметров его величина равна около 41 %.

Отсюда видно, что существует достаточно большой резерв для повышения эффективности МПСЭ при условии получения прямозонных полупроводников с лучшим соответствием требуемых ширин запрещенных зон по мере увеличения числа ячеек $n$.

Полученные зависимости, с одной стороны, определяются конкретным выбором ячеек МПСЭ. С другой стороны, в условиях AM1.5 они в значительной степени зависят от сильной немонотонности спектра солнечного излучения.

Как показало сравнение полученных в работе результатов с результатами, приведенными в работах других авторов, между ними имеется достаточно хорошее соответствие.

В настоящей работе, в отличие от работы [5], повышение эффективности МПСЭ достигается не за счет многократного отражения и перепоглощения света, а за счет улучшения условий теплоотвода. Преимущества подхода работы [5] могут быть реализованы тогда, когда рекомбинация Шокли – Рида несущественна по сравнению с излучательной рекомбинацией. Преимущества подхода, описанного в настоящей работе, реализуются и в случае, когда рекомбинация Шокли-Рида доминирует по сравнению с излучательной рекомбинацией.

Развитый подход может быть использован для оптимизации параметров МПСЭ.

Литература

[1] http://www.nrel.gov/

[2] P.T. Landsberg, V. Badescu. Progress in Quantum Electronics. Volume 22, Issue 4, July 1998, Pages 211–230.

[3] P.T. Landsberg, V. Badescu. Solar energy conversion: list of efficiencies and some theoretical considerations Part II—Results // Progress in Quantum Electronics Volume 22, Issue 4, July 1998, Pages 231–255.

[4] M.A. Green. Third generation photovoltaic. Springer. 160 p.

[5] D. Ding, S.R. Johnson, S.-Q. Yu, S.-N. Wu and Y.-H. Zhang. J. Appl. Phys., **110,** 123104 (2011).

[6] Ж.И.Алферов, В.М. Андреев, В.Д. Румянцев. ФТП, **38** (8)**,** 937 (2004).

[7] M.F. Dias-Aquado, J. Grinbaum, W.T. Fowler and E.G. Lightsey. Small satellite-thermal design, tes, and analysis. Proceedings of the SPIE, **6221**, 622109 (2006)

[8] S.M. Sze and Kwok K. Ng, Physics of Semiconductor Devices, $3^{rd}$ Ed. (John Wiley and Sons, 2007).

[9] В.М. Андреев, В.В. Евстропов, В.С. Калиновский, В.М. Лантратов, В.П Хвостиков. ФТП, **43** (5), 671 (2009).

[10] A.P.Gorban, A.V.Sachenko, V.P.Kostylyov and N.A. Prima. Semiconductor Physics, Quantum Electronics and Optoelectronics, , **3** (3), 322 (2000).

[11] O.Yu. Borkovskaya, N.L. Dmitruk, V.G. Lyapin, A.V. Sachenko. Thin Solid Films, **451-452**, 402 (2004).

12. http://matprop.ru

13. М.М. Колтун. Селективные оптические поверхности преобразователей солнечной энергии (М., Наука, 1979. 215 с.)

[14] M.A. Green, K. Emery, Y. Hishikawa, W. Warta and E.D. Dunlop, Prog. Photovolt: Res. Appl. **20**, 12 (2012).

[15] R.R. King, D.C. Law, K.M. Edmondson, C.M. Fetzer, G.S. Kinsey, H. Yoon, D.D. Krut, J.H. Ermer, R.A. Sherif and N.H. Karam, Advances in OptoElectronics **2007**, 29523 (2007).

[16] B. Mitchell, G. Peharz, G. Siefer, M. Peters, T. Gandy, J.C. Goldschmidt, J. Benick, S.W. Glunz, A.W. Bett and F. Dimroth, Prog. Photovolt: Res. Appl. **19**, 61 (2011).

[17] Р.И. Джиоев, К.В. Кавокин. ФТТ, **33**(10), 2228 (1991)

[18] A. Barnett, D. Kirkpatrick, C. Honsberg, D. Moore, M. Wanlass, K. Emery, R. Schwartz, D. Carlson, S. Bowden1, D. Aiken, A. Gray, S. Kurtz, L. Kazmerski, T. Moriarty, M. Steiner, J. Gray, T. Davenport, R. Buelow, L. Takacs, N. Shatz, J. Bortz, O. Jani1, K. Goossen, F. Kiamilev, A. Doolittle, I. Ferguson, B. Unger, G. Schmidt, E. Christensen, D. Salzman. 22$^{nd}$ European photovoltaic solar energy conference. Milan, Italy, 2007. P. 1-6.